\documentclass[final,5p,times,twocolumn]{elsarticle}
\journal{Physics Letters B}
\usepackage{amssymb,amsmath}
\usepackage{slashed}
\allowdisplaybreaks[3]
\usepackage{calrsfs}
\DeclareMathAlphabet{\pazocal}{OMS}{zplm}{m}{n}
\usepackage[dvipdfmx]{hyperref}

\begin{document}

\begin{frontmatter}
\title{
Equation-of-motion and Lorentz-invariance relations\\
for tensor-polarized parton distribution functions of spin-1 hadrons
}
\author[1,2]{S. Kumano}
\ead{shunzo.kumano@kek.jp}
\author[3,4]{Qin-Tao Song}
\ead{songqintao@zzu.edu.cn}
\address[1]{KEK Theory Center,
             Institute of Particle and Nuclear Studies, 
             High Energy Accelerator Research Organization (KEK),\\
             Oho 1-1, Tsukuba, Ibaraki, 305-0801, Japan}
\address[2]{J-PARC Branch, KEK Theory Center,
             Institute of Particle and Nuclear Studies, KEK, \\
           and Theory Group, Particle and Nuclear Physics Division, 
           J-PARC Center, 
           Shirakata 203-1, Tokai, Ibaraki, 319-1106, Japan}
\address[3]{School of Physics and Microelectronics, Zhengzhou University, 
             Zhengzhou, Henan 450001, China}
\address[4]{
CPHT, CNRS, Ecole Polytechnique, Institut Polytechnique de Paris,
Route de Saclay, 91128 Palaiseau, France
}
\begin{abstract}
Structure functions of polarized spin-1 hadrons will 
be measured at various accelerator facilities in the near future.
Recently, transverse-momentum-dependent and collinear 
parton distribution functions were theoretically proposed 
at twist 3 and twist 4 in addition to the twist-2 ones, 
so that full investigations became 
possible for structure functions of spin-1 hadrons 
in the same level with those of the spin-1/2 nucleons.
Furthermore, twist-3 tensor-polarized multiparton distribution functions
were also recently found for spin-1 hadrons.
In this work, we show relations among the collinear parton- 
and multiparton-distribution functions for spin-1 hadrons by using 
equation of motion for quarks.
These relations are valuable in constraining the distribution functions
and learning about multiparton correlations in spin-1 hadrons.
\end{abstract}

\end{frontmatter}

\date{December 25, 2021}

\section{Introduction}
\label{introduction}

High-energy spin physics has been an exciting field in physics 
since late 1980's for finding the origin of nucleon spin.
Now, the longitudinally polarized parton distribution functions (PDFs)
are relatively well determined except for the gluon distribution.
Partonic angular-momentum contributions are not still determined
experimentally, and their studies are in progress with measurements
of generalized parton distributions (GPDs) \cite{gpds}.
In addition, the field of finding the origin of hadron masses
is also rapidly progressing because similar theoretical formalisms 
are used and the same GPDs or generalized distribution amplitudes
(timelike GPDs) can be used for probing quark and 
gluon composition of hadron masses \cite{kst-2018}. 

On the other hand, structure functions of spin-1 hadrons are quite 
interesting in providing different aspects of hadron polarizations 
because of the existence of new tensor-polarization observables, 
which do not exist in the spin-1/2 nucleons.
As for the tensor-polarized structure functions, 
there exist four functions $b_{1-4}$ in charged-lepton 
deep inelastic scattering from a spin-1 target
such as the deuteron \cite{spin-1-deuteron-sfs},
and there was a measurement on $b_1$ \cite{Airapetian:2005cb}.
There were theoretical studies on the spin-1 physics
on the $b_1$ sum rule \cite{b1-sum},
its second moment \cite{b1x-sum}, 
fragmentation functions \cite{Ji-1994},
tensor-polarized PDFs in the proton-deuteron Drell-Yan process
\cite{pd-drell-yan,Kumano:2016ude},
GPDs \cite{gpds-spin-1},
projection operators on $b_{1-4}$ \cite{b1-4-projections},
optimum tensor-polarized PDFs \cite{tensor-pdfs},
standard-deuteron model prediction for $b_1$ 
\cite{b1-convolution,tagged-ed},
effects of pions and hidden-color state in $b_1$ \cite{miller-b1},
and gluon transversity 
\cite{gluon-trans-1,gluon-trans-2,ks-trans-g-2020}.
In addition to twist 2 \cite{Bacchetta-2000}, 
transverse-momentum-dependent parton distribution functions (TMDs), 
PDFs, and their sum rules were investigated at twist 3 
and twist 4 \cite{ks-tmd-2021}, recently.
Furthermore, 
a twist-2 relation and a sum rule
were obtained for the twist-3 function $f_{LT}$
with investigations
on possible twist-3 multiparton distributions 
\cite{ks-ww-bc-2021}.

In future, there are experimental projects \cite{spin-1-exp}
to investigate structure functions of the spin-1 deuteron at the
Thomas Jefferson National Accelerator Facility (JLab), 
Fermilab (Fermi National Accelerator Laboratory), 
Nuclotron-based Ion Collider fAcility (NICA),
LHC (Large Hadron Collider)-spin,
and electron-ion colliders (EIC, EicC). 
Therefore, the field of spin-1 hadrons could become 
an exciting topic in hadron physics
for investigating exotic aspects of hadrons and nuclei,
possibly beyond the simple
bound systems of nucleons for the deuteron.

In this work, we show useful relations among
the tensor-polarized PDFs and the multiparton distribution functions 
by using equation of motion for quarks.
Then, a so-called Lorentz-invariance relation is derived
for the tensor-polarized PDFs and 
the multiparton distribution functions.
This kind of studies have been done for structure functions
of spin-1/2 nucleons \cite{eq-motion-lorentz}. Here, we investigate
such relations for tensor-polarized spin-1 hadrons.
In Sec.,\ref{spin-1}, we introduce correlation functions necessary
for explaining possible tensor-polarized PDFs and multiparton 
distribution functions. Then, their relations are obtained 
by using the equation of motion, and a Lorentz-invariance relation is 
derived in Sec.\,\ref{Lorentz-eq-motion}.
Our results are summarized in Sec.\,\ref{summary}.

\section{Correlation functions of spin-1 hadrons}
\label{spin-1}

The PDFs and the multiparton distribution functions
are defined through correlation functions for spin-1 hadrons, 
so that we introduce them in this section.
The correlation function is related to the amplitude to extract 
a parton from a hadron and then to insert it into the hadron
at a different spacetime point $\xi$ as given by
\begin{align}
\ \hspace{-0.40cm}
\Phi_{ij}^{[c]} (k, P, T)  
= \! \! \int \! \frac{d^4 \xi}{(2\pi)^4} \, e^{ i k \cdot \xi}
\langle  P, T \left | \, 
\bar\psi _j (0)  W^{[c]} (0, \xi)  
 \psi _i (\xi)  \, \right | P, T  \rangle , \! \!
\label{eqn:correlation-q}
\end{align} 
where $\psi$ is the quark field,
$k$ is the quark momentum, 
$P$ and $T$ indicate hadron momentum and tensor polarization,
$W^{[c]}$ is the gauge link necessary for the color gauge invariance,
and $c$ indicates the integral path.
In this work, we discuss only the tensor polarization, so that
the spin vector $S$ is not explicitly denoted 
in Eq.\,(\ref{eqn:correlation-q}).
The TMD correlation function
is given by integrating Eq.\,(\ref{eqn:correlation-q})
over the lightcone momentum $k^-$ as
\begin{align}
\Phi^{[c]}_{ij} (x, k_T, P, T ) & =  \int dk^+ dk^- \, 
               \Phi^{[c]}_{ij} (k, P, T ) \, \delta (k^+  -x P^+) ,
\label{eqn:correlation-tmd}
\end{align}
where $x$ is the momentum fraction carried by a parton
as defined by $k^+ =x P^+$, and $k_T$ is the transverse momentum.
The lightcone vectors $n$ and $\bar n$ defined as
\begin{align}
n^\mu =\frac{1}{\sqrt{2}} (\, 1,\, 0,\, 0,\,  -1 \, ), \ \ 
\bar n^\mu =\frac{1}{\sqrt{2}} (\, 1,\, 0,\, 0,\,  1 \, ) ,
\label{eqn:lightcone-n-nbar}
\end{align} 
are used in this paper.

Furthermore, if the function is integrated over the transverse momentum,
we obtain the collinear correlation function as
\begin{align}
& \ \hspace{-0.31cm}
\Phi_{ij} (x, P, T ) 
  = \int d^2 k_T \, \Phi^{[c]}_{ij} (x, k_T, P, T )
\nonumber \\
& \ \hspace{-0.49cm}
= \! \int  \frac{d\xi^-}{2\pi} \, e^{ixP^+ \xi^-} \!
\langle \, P , T \left | \, 
\bar\psi _j (0) \,  W (0, \xi \, |\, n)  \,
\psi _i (\xi)  \, \right |  P, \,  T \,
\rangle _{\xi^+ =0, \, \vec\xi_T=0} .
\label{eqn:correlation-pdf}
\end{align}
In investigating various polarized collinear distribution functions, 
it is useful to define the $k_T$-weighted collinear 
correlation function by \cite{Boer:2003bw,br-book}
\begin{align}
( \Phi_\partial^{[\pm]\, \mu} ) _{ij} (x, P, T ) 
  = \int d^2 k_T \,   k_T^{\, \mu}
\, \Phi^{[\pm]}_{ij} (x, k_T, P, T ) .
\label{eqn:kt-weighted-correlation}
\end{align}
Although some collinear distribution functions vanish
due to the time-reversal invariance, 
the $k_T$-weighted distributions could exist.
The superscript index $c=\pm$ indicates 
the direction of the integral path, namely
the plus or minus direction of the coordinate $\xi^-$ ($n^-$)
as shown in Fig.\,4 of Ref.\,\cite{ks-tmd-2021}.
For example, the sign $+$ and $-$ are associated 
with the correlation functions in the simi-inclusive 
deep inelatic scattering (SIDIS) and the Drell-Yan process,
respectively \cite{ks-tmd-2021}.
For discussing twist-3 PDFs, multiparton (three-parton in this work) 
correlation functions are defined with the gluon field tensor 
$G^{+ \mu}$, the gluon field $A^\mu$, 
or the covariant derivative $D^\mu$ as
\begin{align}
(\Phi_X^{\,\mu})_{ij} (x_1, x_2, P, T)
& 
= \int  \! \frac{d \xi_1^{\,-}}{2\pi}   \frac{d \xi_2^{\,-}}{2\pi}  
 \,    e^{i x_1 P^+ \xi_1^{\,-}}  e^{i (x_2-x_1) P^+ \xi_2^{\,-}} 
\nonumber \\
& \ \hspace{0.50cm} \times
\langle \, P, T \left | \,  \bar\psi _j (0) \, 
Y^{\mu}( \xi_2^{\,-} ) 
\, \psi _i (\xi_1^{\,-})  \,
  \right | P, T \, \rangle ,
\nonumber \\
& \ \hspace{-1.00cm} 
X \, (Y^{\mu}) 
= G \, (g \, G^{+ \mu}), \ \ A \, (g \, A^\mu), \ \ D \, (i D^\mu), 
\label{eqn:3parton-correlation}
\end{align} 
where the gauge link is not explicitly written and
$g$ is the QCD coupling constant.
The field tensor $G^{\mu \nu}$ is given by the gluon field $A^\mu$ as
$G^{\mu \nu}= \partial^\mu A^\nu - \partial^\nu A^\mu
                  - ig\left[A^{\mu}, A^{\nu} \right]$,
where the color factor is included in the field as
$A^\mu = A^\mu_a \lambda^a/2$ with the Gell-Mann matrix $\lambda^a$.
In this paper, the lightcone gauge ($n \cdot A=A^+ =0$) is used, 
so that the field tensor $G^{+\alpha}$ is expressed by the gluon field
$A^\alpha$ as $G^{+\alpha}=\partial^+ A^\alpha$.
The covariant derivative $D^\mu$ is given by
$D^\mu = \partial^\mu - ig A^\mu$.
These are the correlation functions used in this work
for discussing the tensor-polarized PDFs and 
the multiparton distributions functions.

\section{Relations among PDFs and multiparton distribution\\ 
functions for spin-1 hadrons by equation of motion}
\label{Lorentz-eq-motion}


Using the correlation functions defined in the previous section,
we derive relations among the PDFs and 
the multiparton distribution functions.
Since the correlation functions appear repeatedly in this section,
we abbreviate the momentum $P$ and tensor polarization $T$
hereafter in denoting the correlation functions $\Phi$. 
The transverse gluon field is the only relevant degree of freedom
in the lightcone quantization \cite{ks-ww-bc-2021,brodsky-1998},
so that the transverse index $\alpha \, (=1,2)$ is used
in the following equations 
($A^\alpha = A_T^\alpha$, $\partial^{\,\alpha} = \partial_T^{\,\alpha}$).

In the $k_T$-weighted correlation function of 
Eq.\,(\ref{eqn:kt-weighted-correlation}),
$k_T^{\, \alpha}$ is expressed by the derivative 
$\partial_T^{\, \alpha} =\partial/\partial \xi_{T\alpha}$.
Applying this derivative to the gauge link
and integrating over $\vec k_T$, we express the function 
$\Phi_{\partial}^{\left[\pm \right]\, \alpha}$
by the covariant derivative and the gluon field tensor as 
\cite{Boer:2003bw}
\begin{align}
(\Phi_{\partial}^{\left[\pm \right]\, \alpha})_{ij} (x)
& 
=  \int \frac{d\xi^{\,-}}{2 \pi} \, e^{i xP^+ \xi^{\,-}} 
    \, \bigg[ \, \langle \, P, T \, | 
     \, \bar{\psi}_j(0) \, i D^{\alpha} \, 
        \psi_i(\xi^-) \, | \, P, T \, \rangle  
\nonumber \\
& \ \hspace{-0.10cm}
  - \langle \, P, T \, | \, \bar{\psi}_j(0) \! 
  \int_{\pm \infty}^{\,\xi^{\,-}} \! d\eta^- g G^{+ \alpha}(\eta^-) \,
  \psi_i(\xi^{\,-}) \, 
  | \, P, T \, \rangle \,  \bigg] 
\nonumber \\
& \ \hspace{-1.00cm}
= \int \frac{d\xi^-}{2 \pi} \, e^{i xP^+ \xi^{\,-}}
  \, \bigg[ \, \langle \, P, T \, | \, \bar{\psi}_j(0) \,
          i \partial^{\alpha} \,
          \psi_i(\xi^{\,-}) \, | \, P, T \, \rangle  
\nonumber \\
& \ \hspace{-0.10cm}
  +   \langle \, P, T \, | \, \bar{\psi}_j(0) 
         \, g \, A^{\alpha}
         (
         \xi^{\,-}
         =\pm \infty) \,
          \psi_i(\xi^{\,-}) \, | \, P, T \, \rangle  \, \bigg] .
\label{eqn:kt-co}
\end{align}
The second equation is obtained by using 
$ G^{+ \alpha}(\eta^-) = \partial_- A^\alpha (\eta^-) $.
The field-tensor relation $G^{+\alpha}=\partial^+ A^\alpha$
is inverted as
\cite{Boer:1997bw}
\begin{align}
A^\alpha (\xi^{\,-}) 
& = \frac{A^\alpha (\infty) + A^\alpha (-\infty)}{2}
- \frac{1}{2} \int_{-\infty}^\infty 
\! d\eta^- 
\epsilon (\eta^- - \xi^{\,-}) \, G^{+\alpha} (\eta^-) ,
\label{eqn:A-alpha}
\end{align}
\vspace{-0.30cm}

\noindent
where $\epsilon (x)$ is the sign function \cite{ks-ww-bc-2021},
by noting the boundary conditions at $\pm \infty$.
Using this relation for Eq.\,(\ref{eqn:kt-co}), we obtain
\begin{align}
\! \!
(\Phi_{\partial}^{\left[\pm \right]\, \alpha})_{ij} (x)
& =\int \frac{d\xi^{\,-}}{2 \pi} \, e^{ix P^+ \xi^{\,-}} \,
   \bigg[ \, \langle \, P, T \, | \, \bar{\psi}_j(0) \, i \partial^{\alpha} \,
           \psi_i(\xi^{\,-}) \, | \, P, T \, \rangle  
\nonumber \\
& \ \ \ \hspace{0.00cm}
      + \langle \, P, T \, | \, \bar{\psi}_j(0) \, 
            \, g \,  \bigg\{   \frac{  A^{\alpha}( \infty) 
                        + A^{\alpha}(- \infty)}{2} 
\nonumber \\
& \ \ \ \hspace{0.90cm}     
       \pm  \frac{  A^{\alpha}( \infty)
           -A^{\alpha}(- \infty)}{2}  \bigg\} \, \psi_i(\xi^{\,-}) \, 
        | \, P, T \, \rangle  \, \bigg]  .
\label{eqn:kt-co1}
\end{align}
Here, the first and second lines are T-even terms
which are expressed by T-even $k_T$-weighted PDFs, 
whereas the third line is a T-odd term
expressed by T-odd distributions \cite{Boer:1997bw}.
The $k_T$-weighted PDFs are defined in Eq.\,(\ref{eqn:trans-moments-TMDs}).
The T-odd $k_T$-weighted PDFs exist although ordinary T-odd PDFs
should vanish due to the time-reversal invariance.
The T-odd distributions are actually gluonic-pole effects 
which are reflected as the difference between
$A^{\alpha}_T( \infty)$ and 
$A^{\alpha}_T(- \infty)$ 
\cite{Boer:2003bw}. 
If one considers 
$\Phi_{\partial}^{\left[+ \right] \, \alpha} (x) 
 +\Phi_{\partial}^{\left[- \right] \, \alpha} (x)$, 
the T-odd distributions vanish,
which is related to 
the TMD relation 
$f_{\text{SIDIS}}(x, k_T^2)=-f_{\text{Drell-Yan}}(x, k_T^2)$
between the TMDs of the SIDIS
and the Drell-Yan process.
If the boundary condition were imposed as 
$A^\alpha (\infty) = A^\alpha (-\infty) $,
the T-odd term vanishes in Eq.\,(\ref{eqn:kt-co1})
\cite{Boer:2003bw}.

Next, we try to obtain the $k_T$-weighted correlation function
of Eq.\,(\ref{eqn:kt-weighted-correlation})
in terms of transverse-momentum moments of TMDs.
First, the twist-2 TMD correlation function is given as 
\cite{Bacchetta-2000,ks-tmd-2021}
\begin{align}
& \Phi^{[\pm]} (x,k_T)_{\text{twist-2}} = 
\, \frac{1}{2} \, \bigg [ 
- f_{1LT}^{[\pm]} (x,k_T^2) \frac{k_T \cdot S_{LT}}{M} {\slashed{\bar n}}
\nonumber \\
&
+ g_{1LT}^{[\pm]} (x,k_T^2) \frac{\epsilon_{T}^{\alpha\mu} S_{LT\alpha} k_{T\mu}}{M}
              \gamma_5 {\slashed{\bar n}}
+ h_{1LL}^{\perp [\pm]} (x,k_T^2)
 S_{LL} \frac{k_{T\alpha} \sigma^{\,\alpha\mu}}{M} 
              \bar n_\mu
\nonumber \\
&
- h_{1TT}^{\prime [\pm]} (x,k_T^2)
 \frac{k_{T\alpha} S_{TT}^{\alpha \beta} 
        \sigma_{\beta\mu}}{M} \bar n^\mu
+ h_{1TT}^{\perp [\pm]} (x,k_T^2) 
 \frac{k_{T\alpha} 
           S_{TT}^{\alpha \beta} k_{T\beta}}{M^2}
           \frac{k_{T\rho} \sigma^{\,\rho \mu}}{M} \bar n_\mu
\nonumber \\
&
+ \text{$k_T$-even terms} \, \bigg ] ,
\label{eqn:TMD-correlation}
\end{align}
where $M$ is the mass of a spin-1 hadron.
The path dependence $[\pm]$ of the TMDs is often
abbreviated, for example, as $f_{1LT} (x,k_T^2)$
instead of $f_{1LT}^{[\pm]} (x,k_T^2)$. However,
we keep it here to show path-dependent
relations in Eq.\,(\ref{eqn:T-odd-even}).
By the $k_T$-weighted integral, the $ h_{1TT}^\perp$ vanishes
because of $S_{TT}^{11}=-S_{TT}^{22}=-1$.
From Eq.\,(\ref{eqn:TMD-correlation}),
the $k_T$-weighted correlation function is expressed by 
transverse-momentum moments 
of the remaining four TMDs as
\cite{Bacchetta-2000,ks-tmd-2021}
\begin{align}
\Phi_{\partial}^{\, [\pm] \, \alpha} (x) 
& = \frac{M}{2} 
    \bigg[  f_{1LT}^{\, [\pm] \, (1)} (x) \, S_{LT}^{\alpha} \,
    \slashed{\bar n} 
    + g_{1LT}^{[\pm] \, (1)} (x) \, 
\epsilon_{T}^{\alpha\mu} S_{LT \mu} \, \gamma_5 \, \slashed{\bar n} 
\nonumber \\
& \ \hspace{-0.25cm}
- h_{1LL}^{\perp \, [\pm] \,  (1)} (x) 
  S_{LL} \sigma^{\alpha \mu} \bar n_{\mu}
+ h_{1TT}^{\prime \, [\pm] \,  (1)} (x)
 \, S_{TT}^{\alpha \beta} \,
        \sigma_{\beta \mu} \, \bar n^{\,\mu}  \bigg] .
\label{eqn:kt-co2}
\end{align}
Here, only the leading-twist $k_T$-weighted PDFs are included
by neglecting higher-twist PDFs, and
the transverse-momentum moments of the TMDs are given by
\begin{align}
f^{\, (1)} (x) = \int \! d^2 k_T \frac{\vec k_T^{\,2}}{2 M^2} \, f(x,k_T^2) .
\label{eqn:trans-moments-TMDs}
\end{align}
The only T-even distribution is $f_{1LT}^{\,(1)}$ and the others are
T-odd ones \cite{ks-tmd-2021} to satisfy 
\begin{alignat}{3}
f_{1LT}^{\, [+] \, (1)} (x) & = f_{1LT}^{\, [-] \, (1)} (x), \ \ \ &
g_{1LT}^{[+] \, (1)}  (x) & = - g_{1LT}^{[-] \, (1)} (x), \ \ 
\nonumber \\
h_{1LL}^{\perp \, [+] \, (1)}  (x) 
       & = - h_{1LL}^{\perp \, [-] \, (1)} (x) , \ \ \ &
h_{1TT}^{\prime \, [+] \, (1)} (x)  
       & = - h_{1TT}^{\prime \, [-] \, (1)} (x) .
\label{eqn:T-odd-even}
\end{alignat} 

By using the gluon-field expression in Eq.\,(\ref{eqn:A-alpha})
and the sign function $\epsilon (x)$ in 
Eq.\,(3.36) of Ref.\,\cite{ks-ww-bc-2021}, 
the multiparton correlation function $\Phi_A^\alpha (x,y)$ 
in Eq.\,(\ref{eqn:3parton-correlation}) 
is related to another one $\Phi_G^\alpha (x,y)$ as
\cite{Boer:1997bw}
\begin{align}
\Phi_A^{\,\alpha} (x, y)
& =\delta(x-y) \, \frac{\Phi_{A(- \infty)}^{\,\alpha} (x)
                          + \Phi_{A( \infty)}^{\,\alpha} (x) }{2}  
\nonumber \\
& \ \ \ \ 
- {\cal P} \left[ \frac{i}{P^+(x-y)} \right ] 
  \Phi_G^\alpha (x,y) ,
\label{eqn:a-g-t}
\end{align} 
where ${\cal P}$ indicates the principal integral.
Then, from Eqs.\,(\ref{eqn:3parton-correlation}),
(\ref{eqn:kt-co1}), and (\ref{eqn:a-g-t}), 
the multiparton correlation function defined with 
the covariant derivative 
$\Phi_D^\alpha(x, y)$ is given by
\begin{align}
\Phi_D^\alpha(x, y)
=\delta(x-y)  
\frac{1}{P^+}
\tilde{\Phi}^{\alpha}(x)  
- {\cal P} \left [ \frac{i}{P^+(x-y)} \right ]
  \Phi_G^\alpha(x, y) ,
\label{eqn:a-g-t1}
\end{align} 
where $\tilde{\Phi}^{\alpha}$ is defined by
the average of $\Phi_{\partial}^{[+] \,\alpha}$ and 
$\Phi_{\partial}^{[-] \,\alpha}$ as
\begin{align}
\! \hspace{-0.20cm}
\tilde{\Phi}^{\alpha}_{ij} (x)
& 
\equiv \frac{( \Phi_{\partial}^{\left[+ \right] \,\alpha})_{ij} (x) 
+ (\Phi_{\partial}^{\left[- \right] \,\alpha})_{ij} (x)}{2} 
\nonumber \\
& 
=\int \frac{d\xi^-}{2 \pi} \, e^{i xP^+ \xi^-} 
\bigg [ \, \langle \, P, T \, | \, \bar{\psi}_j(0) \, i \partial^{\alpha} \,
     \psi_i(\xi^-) \, | \, P, T \, \rangle 
\nonumber \\
& \ \hspace{+0.1cm}
 + \langle \, P, T \, | \, \bar{\psi}_j(0) \, g \frac{  A^{\alpha}( \infty) \,
 + A^{\alpha}(- \infty)}{2}   \psi_i(\xi^-) \, | \, P, T \, \rangle  \bigg ] . 
\label{eqn:kt-co5}
\end{align}
This correlation function $\tilde{\Phi}^{\,\alpha}$ is 
the T-even part of Eq.\,(\ref{eqn:kt-co1}).
Using the $k_T$-weighted correlation function in Eq.\,(\ref{eqn:kt-co2}),
we find that it is given by the T-even function $f_{1LT}^{\,(1)} (x)$ 
as
\begin{align}
\tilde{\Phi}^{\, \alpha} (x) 
= \frac{M}{2} f_{1LT}^{\, (1)} (x)
\, S_{LT}^\alpha \, \slashed{\bar n} ,
\label{eqn:t-even-coll}
\end{align}
where 
$f_{1LT}^{\, [+] \, (1)} (x) = f_{1LT}^{\, [-] \, (1)} (x)
 \equiv f_{1LT}^{\, (1)} (x)$.

The multiparton correlation function $\Phi_D^{\,\alpha} (x,y)$
is expressed by the multiparton distribution functions in the same way
with $\Phi_G^{\,\alpha}(x, y)$ of Ref.\,\cite{ks-ww-bc-2021} as 
\begin{align}
\Phi_D^\alpha (x, y)
& = 
\frac{M}{2P^+} 
\bigg[ S_{LT}^\alpha F_{D,LT}(x, y) 
+ i \epsilon_{\bot}^{\alpha \mu} S_{LT \mu} \gamma_5  G_{D,LT}(x, y)  
\nonumber \\
& \ \hspace{-0.60cm}
+ S_{LL}
\gamma^{\alpha} H_{D,LL}^\perp (x, y)
+ S_{TT}^{\alpha \mu} \gamma_{\mu}  H_{D,TT} (x, y)  \bigg]
\, \slashed{\bar n}  .
\label{eqn:agluond}
\end{align} 
Substituting Eqs.\,(\ref{eqn:t-even-coll}), (\ref{eqn:agluond}),
and 
$\Phi_G^{\,\alpha}(x, y)$ in Eq.\,(3.32)
of Ref.\,\cite{ks-ww-bc-2021}
into Eq.\,(\ref{eqn:a-g-t1}), we obtain
the relations between the multiparton distribution functions as
\begin{align}
F_{D,LT} (x, y)
& = \delta(x-y) f_{1LT}^{\,(1)} (x)
  + {\cal P} \left ( \frac{1}{x-y} \right) F_{G,LT}(x, y) ,
\nonumber \\
G_{D,LT} (x, y) & = {\cal P} \left( \frac{1}{x-y} \right) 
                          G_{G,LT} (x, y) ,
\nonumber \\
H_{D,LL}^\perp (x, y) & = {\cal P} \left( \frac{1}{x-y} \right) 
                          H_{G,LL}^\perp (x, y) ,
\nonumber \\
H_{D,TT} (x, y) & = {\cal P} \left( \frac{1}{x-y} \right) 
H_{G,TT} (x, y) .
\label{eqn:agluond1}
\end{align}
Similar relations are given for the multiparton distribution
functions of the nucleons in Refs.\,\cite{Eguchi:2006qz} and 
\cite{Zhou:2009jm}.
From these relations, Eq.\,(\ref{eqn:T-odd-even}),
and Eq.\,(3.33) of Ref.\,\cite{ks-ww-bc-2021},
we find that the function $F_{D,LT}(x, y)$ is even
under the exchange of $x$ and $y$ and the other functions
[$G_{D,LT}(x, y)$, $H_{D,LL}^\perp(x, y)$, $H_{D,TT}(x, y)$]
are odd as
\begin{alignat}{2}
F_{D,LT}(x, y) & =   F_{D,LT}(y, x), \ \ &
G_{D,LT}(x, y) & = - G_{D,LT}(y, x),
\nonumber \\
H_{D,LL}^\perp(x, y) & = - H_{D,LL}^\perp(y, x), \ \ &
H_{D,TT}(x, y)       & = - H_{D,TT}(y, x).  
\label{eqn:time-rev-PD}
\end{alignat} 


The equation of motion for quarks is $(i\slashed {D} -m) \psi=0$,
which is then multiplied by 
$i \hspace{0.02cm} \sigma^{+ \alpha}$ 
and written by the lightcone coordinates as
\begin{align}
i(\gamma^+ D^\alpha -\gamma^\alpha D^+) \psi
& +i \epsilon_T^{\alpha \mu} \gamma_{\mu} \gamma_{5}  i D^{+} \psi 
\nonumber \\
& 
-i \epsilon_T^{\alpha \mu} \gamma^{+}  \gamma_{5}  i D_{\mu} \psi  
+i \, m \, \sigma^{+ \alpha} 
\psi=0 ,
\label{eqn:eom}
\end{align}
where $m$ is a quark mass.
The constraint from this equation of motion is given
by traces of collinear correlation functions 
$\Phi_D^{\, \mu \, (= \, \alpha,+)} (x)$ as
\begin{align}
& 
\int  \frac{d\xi^-}{2\pi} \, e^{ixP^+ \xi^-}
\langle \, P , T \, \big | \, 
\bar\psi (0) \, \big [ \,
i(\gamma^+ D^\alpha -\gamma^\alpha D^+) 
+i \epsilon_T^{\alpha \mu} \gamma_{\mu} \gamma_{5}  i D^{+} 
\nonumber \\
& \ \hspace{2.0cm}
-i \epsilon_T^{\alpha \mu} \gamma^{+}  \gamma_{5}  i D_{\mu} 
+i \, m \, \sigma^{+ \alpha} \, \big ] \,
\psi  (\xi^{\,-})  \, \big | \, P, \,  T \, \rangle 
\nonumber \\
& \ 
= P^+ \text{Tr} \, [\Phi_D^\alpha (x) \, \gamma^+]
- P^+ \text{Tr} \, [\Phi_D^+ (x) \, \gamma^\alpha]
+ P^+  i \epsilon_T^{\alpha\mu} \text{Tr} \, [\Phi_D^+ (x) \, \gamma_\mu \gamma_5]
\nonumber \\
& \ \hspace{2.0cm}
- P^+ i \epsilon_T^{\alpha\mu} \text{Tr} \, [\Phi_{D\mu} (x) \, \gamma^+ \gamma_5]
+ im \text{Tr} \, [\Phi (x) \, \sigma^{+\alpha}]
\nonumber \\
& \ 
= 0 .
\label{eqn:eom-correlation-1}
\end{align}
Here, the collinear correlation function $\Phi_D^{\, \mu} (x)$ 
is defined by integrating the correlation function 
$\Phi_D^{\,\mu} (y,x)$ over the variable $y$ as
\begin{align}
\Phi_D^{\,\mu} (x) \equiv \int_{-1}^1 
dy \, \Phi_D^{\,\mu} (y,x).
\label{eqn:PhiD-integ}
\end{align}
It should be noted that only the transverse ($\mu=\alpha$)
correlation functions are associated with the multiparton
distribution functions by Eq.\,(\ref{eqn:agluond}).
In expressing integrals of multiparton distribution functions over $y$
in the following [Eqs.\,(\ref{eqn:eom1}), (\ref{eqn:eom1d}),
(\ref{eqn:lir2}), (\ref{eqn:eom2b})],
the variables $x$ and $y$ are interchanged by using 
Eq.\,(\ref{eqn:time-rev-PD}).
From the definition of Eq.\,(\ref{eqn:3parton-correlation}),
the correlation function $\Phi_D^+ (x,y)$ 
is expressed by the collinear correlation function $\Phi (x)$ as
\begin{align}
\Phi_D^+ (x,y) = \delta (x -y) \, x \, \Phi (x) ,
\label{eqn:PhiD+}
\end{align}
where $\Phi (x)$ is given by \cite{ks-ww-bc-2021}
\begin{align}
\Phi (x) 
& = \frac{1}{2} \bigg [ \,
S_{LL} \, \slashed{\bar n} \, f_{1LL} (x) 
+ \frac{M}{P^+} \, S_{LL} \, e_{LL} (x) 
\nonumber \\
& \ \hspace{0.6cm} 
+ \frac{M}{P^+} \, \slashed{S}_{LT} \, f_{LT} (x) 
+ \frac{M^2}{(P^+)^2} \, S_{LL} \, \slashed{n} \, f_{3LL} (x) \,
\bigg ] .
\label{eqn:collinear-correlation-pdfs}
\end{align}

Calculating Eq.\,(\ref{eqn:eom-correlation-1}) with 
Eqs.\,(\ref{eqn:agluond}), (\ref{eqn:PhiD-integ}),
(\ref{eqn:PhiD+}), and (\ref{eqn:collinear-correlation-pdfs}),
we obtain the relations among the tensor-polarized PDFs
and the multiparton distribution functions as 
\begin{align}
x f_{LT}(x) - \int_{-1}^1 dy \,
\left[ F_{D,LT} (x, y) + G_{D,LT} (x, y) \right]=0 .
\label{eqn:eom1}
\end{align}
A similar relation to this equation was obtained 
in Ref.\,\cite{gluon-trans-2}
by using a different parametrization
for the tensor polarization.
Namely, the collinear twist-3 function $f_{LT}(x)$ is given
by integrating the twist-3 multiparton distribution functions 
$F_{D,LT}$ and $G_{D,LT}$ over one of the momentum-fraction
variables.
Then, this relation is written in terms of 
the multiparton distribution functions defined 
with the field tensor $G^{\mu\nu}$
by using Eq.\,(\ref{eqn:agluond1}) as
\begin{align}
x f_{LT}(x) - f_{1LT}^{\,(1)}(x)
- {\cal P} \int_{-1}^1 dy \,
\frac{F_{G,LT}(x, y) + G_{G,LT} (x, y) }{x-y} = 0 .
\label{eqn:eom1d}
\end{align}
Therefore, the function $f_{LT}(x)$ is also expressed by
the $k_T$-weighted function $f_{1LT}^{\,(1)}(x)$
and other multiparton distribution functions 
$F_{G,LT}$ and $G_{G,LT}$.
Using the relation between the twist-3 structure function
$f_{LT}$ and the twist-2 one $f_{1LL}$ in
Eqs.\,(3.41) and (3.42) of Ref.\,\cite{ks-ww-bc-2021},
and taking the derivative of Eq.\,(\ref{eqn:eom1d})
with respect to $x$, we obtain
\begin{align}
\! \! \! \! \!
\frac{d f_{1LT}^{\,(1)}(x) }{dx} - f_{LT}(x) + \frac{3}{2} f_{1LL}(x)
- 2 {\cal P}   \! \!
 \int_{-1}^1   \!  dy \, \frac{F_{G,LT} (x, y)}{(x-y)^2} =0 . \!
\label{eqn:lir2}
\end{align}
This is a Lorentz-invariance relation for the tensor-polarized
structure functions of spin-1 hadrons in the similar way 
to the ones for the spin-1/2 nucleons \cite{eq-motion-lorentz}.
An equation like Eq.\,(\ref{eqn:lir2}) is conventionally 
called a Lorentz-invariance relation.
Here, the Lorentz invariance means the frame independence of 
twist-3 observables \cite{eq-motion-lorentz}. Therefore,
Lorentz-invariant relations play an important role in twist-3 studies.
Although it is abbreviated in Eq.\,(\ref{eqn:correlation-q}), 
there is dependence on the lightcone vector $n$ due to the gauge link
in the correlation function \cite{ks-tmd-2021}.
The Lorentz-invariance relation is affected
by $n$-dependent terms in the correlation function 
as noticed in the papers of Goeke {\it et al.} (2003) 
and Metz {\it et al.} (2009) in Ref.\,\cite{eq-motion-lorentz}
for the spin-1/2 nucleons. Such effects are included in our formalism.

Multiplying $\gamma^\mu$ on the left-hand side of the Dirac equation and 
using the identity $\gamma^\mu \gamma^\nu = g^{\mu\nu} - i \sigma^{\mu\nu}$,
we obtain $(i D^\mu -i\sigma^{\mu\nu} iD_\nu - m \gamma^\mu )\psi =0$.
Then, taking 
the lightcone component $+$, we have another equation of motion as
\begin{align}
( iD^+ -i  \sigma^{+ \nu} i D_{\nu} - m \gamma^{+}  ) \, \psi=0 .
\label{eqn:eom1a}
\end{align}
The constraint from this equation of motion is written,
in the same way with Eq.\,(\ref{eqn:eom-correlation-1}), as
\begin{align}
& 
\int \frac{d\xi^-}{2\pi} \, e^{ixP^+ \xi^-}
\langle \, P , T \, \big | \, 
\bar\psi (0) \, \big [ 
iD^+ -i  \sigma^{+ \nu} i D_{\nu} - m \gamma^{+} \big ] 
\, \psi  (\xi^{\,-}) \, \big | \, P,  T \, \rangle 
\nonumber \\
& \ \hspace{0.5cm}
= P^+ \text{Tr} \, [\Phi_D^+ (x)]
- P^+ i \, \text{Tr} \, [\Phi_D^+ (x) \, \sigma^{+-}]
\nonumber \\
& \ \hspace{1.3cm}
- P^+ i \, \text{Tr} \, [\Phi_{D\alpha} (x) \, \sigma^{+\alpha}]
- m \text{Tr} \, [\Phi (x) \gamma^+ ]
\nonumber \\
& \ \hspace{0.5cm}
= 0 .
\label{eqn:eom-correlatin}
\end{align}
Then, calculating the traces with Eqs.\,
(\ref{eqn:agluond}), (\ref{eqn:PhiD-integ}),
(\ref{eqn:PhiD+}), and (\ref{eqn:collinear-correlation-pdfs}),
we obtain the integral relation among
the twist-3 PDF $e_{LL}$,
the twist-2 PDF $f_{1LL}$, and
the twist-3 multiparton distribution function $H_{D,LL}^\perp$ as
\begin{align}
x \, e_{LL}(x) - 2 \int_{-1}^1 dy \, H_{D,LL}^\perp (x, y) 
-\frac{m}{M} f_{1LL} (x) =0 .
\label{eqn:eom2a} 
\end{align}
This equation is expressed by another multiparton distribution function 
$H_{G,LL}^\perp$ by using Eq.\,(\ref{eqn:agluond1}) as
\begin{align}
x \, e_{LL}(x)
- 2 {\cal P} \int_{-1}^1 dy \,  \frac{H_{G,LL}^\perp (x, y)}{x-y} 
-\frac{m}{M} f_{1LL} (x) 
=0 .
\label{eqn:eom2b} 
\end{align}
Because of $m/M \ll 1$, the third terms of Eqs.\,(\ref{eqn:eom2a}) 
and (\ref{eqn:eom2b}) could be practically neglected.
Then, the twist-3 functions is described 
only by the multiparton distribution function 
$H_{D,LL}^\perp (x, y)$
or $H_{G,LL}^\perp (x, y)$.
We notice that the distribution $H_{D,LL}^\perp (x, y)$ has 
the corresponding twist-3 PDF $e_{LL}(x)$
by Eq.\,(\ref{eqn:eom2a}).
The distributions $F_{D,LL} (x, y) $ and $G_{D,LT}(x, y)$ 
are related to twist-3 PDF 
$f_{LT}(x)$ 
as shown in Eq. (\ref{eqn:eom1}).
In the nucleon case, all the twist-3 PDFs are expressed 
by twist-3 multiparton distribution functions.
One finds this correspondence in 
Eqs.\,(11), (16), (19), and (20) of 
Ref.\,\cite{Jaffe:1991ra} in the spin-1/2 case.
However, we find that the corresponding twist-3 PDF 
does not exist for the distribution $H_{D,TT} (x, y)$ 
of a spin-1 hadron in Eq. (\ref{eqn:agluond}).  
This is because the distribution $H_{D,TT}(x, y)$ 
is associated with the tensor-polarization parameter 
$S_{TT}^{\mu \mu}$, and there is no twist-3 PDF which 
is associated with the parameter $S_{TT}^{\mu \mu}$ 
\cite{ks-trans-g-2020,ks-tmd-2021,ks-ww-bc-2021}.
In this case, no twist-3 PDF is related to 
$H_{D,TT} (x, y)$ by the equation of motion.

\section{Summary}
\label{summary}

From the equation of motion for quarks,
we derived relations among the tensor-polarized
distribution functions and twist-3 multiparton distribution functions
defined by the field tensor.
We found the relations from
the equation of motion for quarks as
\begin{align}
& 
x f_{LT}(x) - f_{1LT}^{\,(1)}(x)
- {\cal P} \int_{-1}^1 dy \,
\frac{F_{G,LT}(x, y) + G_{G,LT} (x, y) }{x-y} = 0 ,
\nonumber \\
&
x \, e_{LL}(x) - 2 {\cal P} \int_{-1}^1 dy \,  \frac{H_{G,LL}^\perp (x, y)}{x-y} 
-\frac{m}{M} f_{1LL} (x) 
=0 ,
\nonumber 
\end{align}
for the twist-3 PDF $f_{LT}$,
the trasverse-momentum moment PDF $f_{1LT}^{\,(1)}$, and 
the multiparton distribution functions $F_{G,LT}$ and $G_{G,LT}$;
for the twist-3 PDF $e_{LL}$, the twist-2 PDF $f_{1LL}$,
and the multiparton distribution function $H_{G,LL}^\perp$.
Then, the Lorentz-invariance relation was obtained as
\begin{align}
\frac{d f_{1LT}^{\,(1)}(x) }{dx}
- f_{LT}(x) + \frac{3}{2} f_{1LL}(x)
- 2 {\cal P}   \! 
 \int_{-1}^1   \!  dy \, \frac{F_{G,LT} (x, y)}{(x-y)^2} =0 .
\nonumber 
\end{align}
These relation are valuable in constraining the tensor-polarized
PDFs and the multiparton distribution functions.

In deriving these relations, we also obtained new relations 
among the multiparton distribution functions defined by 
the field tensor and the covariant derivatives.
First, the function $F_{D,LT} (x, y)$ is expressed
by $f_{1LT}^{\,(1)} (x)$ and $F_{G,LT}(x, y)$ as
\begin{align}
F_{D,LT} (x, y)
 = \delta(x-y) f_{1LT}^{\,(1)} (x)
  + {\cal P} \left ( \frac{1}{x-y} \right) F_{G,LT}(x, y) .
\nonumber 
\end{align}
Next, the functions $G_{D,LT} (x, y)$, $H_{D,LL}^\perp (x, y)$, 
and $H_{D,TT} (x, y)$ are expressed only 
by the corresponding functions defined with the field tensor as
\begin{align}
& G_{D,LT} (x, y) = {\cal P} \left( \frac{1}{x-y} \right) 
                          G_{G,LT} (x, y) ,  
\nonumber \\                         
& 
\text{and same equations for } 
H_{D/G,LL}^\perp (x, y) \text{ and } H_{D/G,TT} (x, y) .
\nonumber 
\end{align}
These studies will be useful for investigating 
the tensor-polarized structure functions of spin-1 hadrons.

\section*{Acknowledgments}

S. Kumano was partially supported by 
Japan Society for the Promotion of Science (JSPS) Grants-in-Aid 
for Scientific Research (KAKENHI) Grant Number 19K03830.
Qin-Tao Song was supported by the National Natural Science Foundation 
of China under Grant Number 12005191, the Academic Improvement Project 
of Zhengzhou University, and the China Scholarship Council 
for visiting Ecole Polytechnique.




\begin{thebibliography}{00}
\bibitem{gpds}
 For introductory review, see 
    M.~Diehl, 
        {Phys. Rep. {388}, 41 (2003)};  
    X.~Ji,  
        {Annu. Rev. Nucl. Part. Sci. {54}, 413 (2004)}.
For recent works, see
    H. Moutarde, P. Sznajder, and J. Wagner, 
       {Eur. Phys. J. C {78} (2018) 890};
    PARTONS project at
       http://partons.cea.fr/partons/doc/html/index.html.
See also
    S. Kumano,  M. Strikman, and K. Sudoh, 
        {Phys. Rev. D {80} (2009) 074003};
    H. Kawamura and S. Kumano,
        {Phys. Rev. D 89 (2014) 054007};
    T. Sawada {\it et al.}, 
        {Phys. Rev. D 93 (2016) 11}.
\bibitem{kst-2018}
   S. Kumano, Qin-Tao Song, and O. V. Teryaev, 
      {Phys. Rev. D 97 (2018) 014020};
   K. Tanaka,  
   {Phys.Rev.D 98 (2018) 034009},
   Y. Hatta, A, Rajan, and K. Tanaka,
   {JHEP 12 (2018) 008};
   K. Tanaka,  
   {JHEP 01 (2019) 120}. 
   For reviews and recent works, see, for example,
   M. V. Polyakov and P. Schweitzer,
      {Int. J. Mod. Phys. A 33 (2018) 1830025};
   X. Ji, 
      {Front. Phys. 16 (2021) 64601}
   and references therein;
    C. Lorc\'{e},
    Eur. Phys. J. C 78 (2018) 120; 
    C. Lorc\'{e}, A. Metz, B. Pasquini, and S. Rodini,
    {JHEP 11 (2021) 121}.
\bibitem{spin-1-deuteron-sfs}
   L. L. Frankfurt and M. I. Strikman, 
      {Nucl. Phys. {A405} (1983) 557};
   P.~Hoodbhoy, R.~L.~Jaffe, and A.~Manohar,
      {Nucl. Phys. B 312 (1989) 571}.
\bibitem{Airapetian:2005cb} 
   A.~Airapetian {\it et al.} (HERMES Collaboration),
      {Phys. Rev. Lett. 95 (2005) 242001}.
\bibitem{b1-sum} 
   F. E. Close and S. Kumano, 
      {Phys. Rev. D 42 (1990) 2377};
   S. Kumano, 
      {J. Phys.: Conf. Series 543 (2014) 012001}.
\bibitem{b1x-sum}
   A. Efremov and O. Teryaev, Sov. J. Nucl. Phys. 36 (1982) 557.
\bibitem{Ji-1994} 
   X. Ji, 
      {Phys. Rev. D {49} (1994) 114}.
\bibitem{pd-drell-yan}
   S. Hino and S. Kumano, 
      {Phys. Rev. D 59 (1999) 094026};
      {D 60 (1999) 054018}.
\bibitem{Kumano:2016ude} 
   S.~Kumano and Qin-Tao Song,
      {Phys. Rev. D 94 (2016) 054022}.
\bibitem{gpds-spin-1} 
   E.~R.~Berger, F.~Cano, M.~Diehl and B.~Pire,
      {Phys. Rev. Lett. {87} (2001) 142302};
   W. Cosyn and B. Pire,
      {Phys. Rev. D {98} (2018) 074020};       
   Bao-Dong Sun and Yu-Bing Dong, 
      {Phys. Rev. D {96} (2017) 036019};
                   {{101} (2020) 096008}.
\bibitem{b1-4-projections} 
   T.-Y. Kimura and S. Kumano,
      {Phys. Rev. D {78} (2008) 117505}.
\bibitem{tensor-pdfs} 
   S. Kumano, 
      {Phys. Rev. D 82 (2010) 017501}.
\bibitem{b1-convolution} 
   H. Khan and P. Hoodbhoy, 
      {Phys. Rev. C {44} (1991) 1219};
   W. Cosyn, Yu-Bing Dong, S. Kumano, and M. Sargsian,
      {Phys. Rev. D 95 (2017) 074036}. 
\bibitem{tagged-ed}
   W. Cosyn and C. Weiss, 
      {Phys. Rev. C {102} (2020) 065204}.
\bibitem{miller-b1} 
   G. A. Miller,  
      {Phys. Rev. C {89} (2014) 045203}.      
\bibitem{gluon-trans-1}
   R. L. Jaffe and A. Manohar, 
      {Phys. Lett. B {223} (1989) 218};
   M.~Nzar and P.~Hoodbhoy,
      {Phys. Rev. D {45} (1992) 2264};
   W. Detmold and P. E. Shanahan, 
      {Phys. Rev. D {94} (2016) 014507};
      Erratum,
                   {{95} (2017) 079902}.
\bibitem{gluon-trans-2}
   J. P. Ma, C. Wang, and G. P. Zhang, 
      {arXiv:1306.6693} (unpublished).
\bibitem{ks-trans-g-2020} 
   S. Kumano, Qin-Tao Song,
      {Phys. Rev. D 101 (2020) 054011};
      {D 101 (2020) 094013}.
\bibitem{Bacchetta-2000} 
   A. Bacchetta and P. Mulders, 
      {Phys. Rev. D 62 (2000) 114004}.
\bibitem{ks-tmd-2021}
   S. Kumano and Qin-Tao Song,
      {Phys. Rev. D 103 (2021) 014025}.
\bibitem{ks-ww-bc-2021}
   S. Kumano and Qin-Tao Song, 
      {JHEP 09 (2021) 141}.
\bibitem{spin-1-exp}
    (JLab) J.-P. Chen {\it et al.},
       {Proposal to Jefferson Lab PAC-38, PR12-11-110 (2011)};
    M. Jones {\it et al.},
       {A Letter of Intent to Jefferson Lab PAC 44, LOI12-16-006 (2016)};
    (Fermilab) D. Geesaman {\it et al.},
       {Proposal to Fermilab PAC, P-1039 (2013)};
       D. Keller {\it et al.}, proposal to be submitted to Fermilab in 2022;
    (NICA) A. Arbuzov {\it et al.},
       {Prog. Nucl. Part. Phys. 119 (2021) 103858};      
    (LHCspin) C. A. Aidala {\it et al.}, 
       {arXiv:1901.08002}; 
    (EIC) R. Abdul Khalek {\it et al.}, 
       {arXiv:2103.05419};
    (EicC) D. P. Anderle {\it et al.}, 
       {Front. Phys. {16} (2021) 64701}.
\bibitem{eq-motion-lorentz}
    For explanations on the Lorentz-invariance and equation-of-motion
    relations, see 
    A. Accardi, A. Bacchetta, W. Melnitchouk, and M. Schlegel,
       {JHEP 11 (2009) 093};
    K.~Kanazawa, Y.~Koike, A.~Metz, D.~Pitonyak, and M.~Schlegel,
       {Phys. Rev. D 93 (2016) 054024}.
    Original and related works are 
    R. D. Tangerman, 
       {Ph.D. thesis, Free University Amsterdam (1996)};
    P. J. Mulders and R. D. Tangerman,
       {Nucl. Phys. B 461 (1996) 197};  
       {Erratum, B 484 (1997) 538};
    D. Boer and P. J. Mulders,
       {Phys. Rev. D 57 (1998) 5780};
    D. Boer, P. J. Mulders, and F. Pijlman,
       {Nucl. Phys. B 667 (2003) 201};  
    K. Goeke, A. Metz, P. V. Pobylitsa, and M. V. Polyakov,
       {Phys. Lett. B 567 (2003) 27};    
    A. V. Belitsky and D. M\"uller, 
        {Nucl. Phys. B 503 (1997) 279};
    A. V. Belitsky,
        {Int. J. Mod. Phys. A 32 (2017) 1730018};
    A. Metz, P. Schweitzer, and T. Teckentrup,
       {Phys. Lett. B {680} (2009) 141};
    A. Rajan, M. Engelhardt, and S. Liuti,
      {Phys. Rev. D {98} (2018) 074022}.  
\bibitem{Boer:2003bw}
   D.~Boer, P.~J.~Mulders, and F.~Pijlman, 
      {Nucl. Phys. B 667 (2003) 201}.
\bibitem{br-book} 
   V. Barone and R. G. Ratcliffe, 
      {{\it Transverse Spin Physics} (World Scientific, Singapore, 2003)}.
\bibitem{brodsky-1998}
   S. J. Brodsky, H.-C. Pauli, and S. S. Pinsky,
      {Phys. Rept. 301 (1998) 299}.   
\bibitem{Boer:1997bw}
   D.~Boer, P.~J.~Mulders, and O.~V.~Teryaev, 
       {Phys. Rev. D 57 (1998) 3057}.
\bibitem{Eguchi:2006qz}
   H.~Eguchi, Y.~Koike, and K.~Tanaka, 
      {Nucl. Phys. B 752 (2006) 1}.
\bibitem{Zhou:2009jm}
   J.~Zhou, F.~Yuan, and Z.~T.~Liang, 
      {Phys. Rev. D 81 (2010) 054008}.
\bibitem{Jaffe:1991ra}
    R.~L.~Jaffe and X.~D.~Ji,
       {Nucl. Phys. B 375 (1992) 527}.
\end{thebibliography}
\end{document}